\documentclass[10pt]{article}

\usepackage{epsfig}

\date{}

\title{
{
\Large \bf Neural Nets and Star/Galaxy Separation in Wide Field
Astronomical Images }}
\author{\large Stefano Andreon$^{1}$, Giorgio Gargiulo$^{2}$, Giuseppe Longo$^{1}$,\\ 
\large Roberto Tagliaferri$^{3}$ and Nicola Capuano$^{2}$ \\
$^{1}${\large Osservatorio Astronomico di Capodimonte, via Moiariello 16,}\\
{\large I-80131 Napoli, Italia, \{andreon, longo\}@na.astro.it}
\\
$^{2}${\large  Universit\`{a} di Salerno, via S. Allende, } 
{\large 84081 Baronissi (SA) Italia}\\
$^{3}${\large DMI, Universit\`{a} di Salerno, INFM unit\`{a} di Salerno, and IIASS}\\
{\large via S. Allende, 84081 Baronissi (SA) Italia, robtag@dia.unisa.it}
}

\begin{document}

\vspace{-2truecm}
\twocolumn
\columnsep 0.8truecm 
\maketitle	 

\thispagestyle{empty}


\begin{center}
{\bf \large Abstract}
\end{center}

{\it
\noindent 
One of the most relevant problems in the extraction of scientifically useful
information from wide field astronomical images (both photographic plates
and CCD frames) is the recognition of the objects against a noisy background
and their classification in unresolved (star-like) and resolved (galaxies)
sources. In this paper we present a neural network based method capable to
perform both tasks and discuss in detail the performance of object
detection in a representative celestial field. The performance of our
method is compared to that of other methodologies often used within the
astronomical community.
}
\\


\begin{center}
{\bf \large 1. Introduction }
\end{center}

\noindent 
Astronomical wide field imaging (hereafter WFI)\ and its most extreme case,
all sky surveys such as the Palomar Sky Surveys (POSS I\ \&\ II), are the
main tools to tackle astronomical problems requiring statistically
significant samples of optically selected objects. In the past, WFI has also
been the main supplier of targets for photometric and spectroscopic
follow-up's at telescopes of the 4 meter class. The exploitation of the new
generation telescopes in the 8 meter class, which are mainly aimed to
observe targets which are too faint to be detected on photographic material
(the POSS-II detection limit in B is $\sim 21.5$ $mag)$ requires new
digitised surveys realized with large format CCD detectors mounted at 2
meter class dedicated telescopes. Much effort is currently devoted worldwide
to construct such facilities: the MEGACAM project at the CFH, the ESO\ Wide
Field Imager at the 2.2 meter telescope, the Sloan - DSS and the ESO/OAC
VST, to quote only some of the ongoing or planned experiments. One aspect
which is never too often stressed is the humongous problem posed by the
handling, processing and archiving of the data produced by these
instruments: the VST, for instance \cite{Arnaboldi98} is expected to produce
a flow of almost 20 GByte of data per night or 10 Tbyte per year of
operation. Such a huge flow of data
cannot be effectively dealt with traditional data reduction packages and
calls for modern A.I. based approaches.
\\
\\
\noindent
In this paper we present a new, neural network (NN) based method, capable to
perform object detection and star/galaxy separation. Due to space limitation
we shall focus our attention mainly on the experimental results relative to
the first step.
\\

\begin{center}
{\bf \large 2. Preprocessing and object detection}
\end{center}

\noindent
After the standard preprocessing of the data \cite{Andreon99} we
perform the following steps:

--  we first run a 3x3 or 5x5 window on the image in order to determine the value of the central pixel;

--  we then use Robust Principal Component Analysis (PCA) NNs to reduce to 3 the dimensionality of the input space.

--  Therefore, since su\-per\-vi\-sed NN's need a large am\-ou\-nt 
of labeled data to obtain a good classi\-fi\-ca\-tion, we use un\-super\-vi\-sed NN's to segment the pixels into six classes (one for the backround and five for the objects).

--  We then group the five objects classes into one and are left with two classes only: background and objects.

--  Finally, in order to split overlapping objects, we run a simple but effective deblending algorithm, capable to isolate the objects against the
noisy background.
\\
\\
\noindent
{\bf 2.1 Preprocessing and object detection}

\medskip
\noindent
PCAs can be neurally realized in various ways; we used a feedforward neural
network with only one layer which is able to extract the principal
components of the stream of input vectors. The structure of the PCA NN can
be summarized as follows: there is one input layer, and one forward layer of
neurons totally connected to the inputs; during the learning phase there are
feedback links among neurons, that classify the network structure as either
hierarchical or symmetric, depending on the feedback connections of the
output layer neurons. Typically, Hebbian type learning rules are used. Many
different versions and extensions of the basic learning algorithm have been
proposed in recent years \cite{Oja82}, \cite{Sanger89}, \cite{Karhunen94}.
After the learning phase, the network becomes purely feedforward. \cite{Karhunen94} proved that PCA neural algorithms can be derived from
optimization problems, such as variance maximization and representation of
error minimization, and derived the so called robust PCA algorithms and
nonlinear PCA algorithms .
More precisely, in the robust generalization of variance maximization, the
objective function $f(z)$ is assumed to be a valid cost function \cite
{Karhunen94} such as $ln\;cos(z)$ and $|z|$. This leads to the adaptation
step of \ the learning algorithm: 
\begin{equation}
w_{ji}^{(t+1)}=w_{ji}^{(t)}+\mu g\left( y_{j}^{(t)}\right) e_{ji}^{(t)}
\label{eq2.5}
\end{equation}

where:

\[
e_{ji}^{(t)}=x_{i}^{(t)}-\sum_{i=1}^{l(j)}y_{i}^{(t)}w_{ji}^{(t)} 
\]

\[
g=\frac{df}{dz} 
\]

In the hierarchical case $l(j)=j$ and in the symmetric case $l(j)=M$. The
learning function $g$, derivative of $f$, is applied separately to each
component of the argument vector. In previous experiments \cite{Tagliaferri
99aass} we found that the hierarchical robust NN of eq.\ref{eq2.5} with
learning function $g^{(t)}=tanh(\alpha x)$ performs better than all the
other PCA NN's and linear PCA.
\\
\\
\noindent {\bf 2.2 Unsupervised NNs}

\medskip
\noindent
The NNs used in this section are based on the classical unsupervised neural
models: Kohonen Self Organizing Maps \cite{Kohonen82}, 
Neural-Gas \cite{Martinetz93}, Growing Cell Structure (GCS) \cite{Fritzke94}, 
on-line K-means clustering algorithm \cite{Lloyd82}, Maximum Entropy NN 
\cite{Rose90}. All these methods allow to partition the input space into
clusters and to assign a weight vector corresponding to the template
characteristic of a cluster in the input space to each neuron. As a
consequence, after the learning, an input pattern is assigned to the class
corresponding to the nearest neuron.
\\
\\
\noindent 
We preferred to reduce the well-known complexity of the post-processing
labeling adding an unsupervised single layer NN to the output of the first
layer NN. In this way, the second layer NN learns from the weights of the
first layer NN and clusters the neurons on the basis of a similarity measure
or a distance. The iteration of this process gives the unsupervised
hierarchical NN's. The number of neurons at each layer decreases from the
first to the output layer, and, as a consequence, the NN takes a pyramidal
aspect. The NN takes as input a pattern $x$ and then the first layer finds
the winner neuron. The second layer takes the first layer winner weight
vector as input and finds the second layer winner neuron and so on until the
top layer. The activation value of the output layer neurons is 1 for the
winner unit and 0 for all the others.
\\
\\
\noindent 
By varying the learning algorithms we obtain different NN's with different
properties and abilities. For instance, by using only SOMs we have a
Multi-layer SOM (ML-SOM) \cite{Koh95} where every layer is a two-dimensional
grid. We can easily obtain \textit{ML-NeuralGas}, \textit{ML-Max\-im\-um Entropy}
or \textit{ML-K} \textit{means} organized on a hierarchy of linear layers 
\cite{tagliaferri99ieee}. The ML-GCS has a more complex architecture and has
at least 3 units for layer. By varying the learning algorithms in the
different layers we can take advantage from the properties of each model
(for example since we cannot have a ML-GCS with 2 output units, then we can
use another NN in the output layer). A hierarchical NN with a number of
output layer neurons equal to the number of the output classes simplifies
the expensive post-processing step of labeling the output neurons in
classes, without reducing the generalization capacity of the NN.\bigskip
\\

\begin{center}
{\bf \large 3. Star/Galaxy separation}
\end{center}

\noindent 
The first step in order to perform star/galaxy separation is to identify the
most significant features. Then we run an optimized \textit{Multi-Layer
Perceptron} (MLP). \cite{Bishop95} and \cite{Press93} summarize methods to
overcome the problems related to local minima and slow time convergence of
the above algorithm.
\\
\\
\noindent 
The object features were chosen following the literature \cite{Jarvis81}, 
\cite{Miller96}, \cite{Odewahn92}, and selected by a simple sequential
forward selection process \cite{Bishop95}, so as to select the most
performing ones. In particular, we took in consideration the following
features:

\begin{itemize}
\item  Six features describing the ellipses circumscribing the objects: the
photometric baricenter coordinates, the isophotal flux, the semimajor axis,
the semiminor axis, the position angle, and the object area ($A$) in pxls.

\item  Twelve features suggested by Odewahn \cite{Odewahn92}: the object
diameter, the ellipticity, the average surface brightness($\left\langle
SuBr\right\rangle $), the central intensity ($I_{0}$), the filling factor,
the area logarithm, the harmonic radius and five simple gradients of the
light distribution $G_{14}$, $G_{13}$, $G_{12}$, $G_{23}$ and $G_{34}$
defined as: 
\[
G_{ij}=\frac{T_{j}-T_{i}}{r_{i}-r_{j}}
\]
\end{itemize}

where $T_{i}$ is the average surface brightness within an ellipse, with
position angle $\alpha $, semimajor axis $r_{i}<a$ and ellipticity $ell$. To
this aim, four equidistant radii $r_{i}$ are selected with $r_{i}=i\,a/4,\ i=1, \ldots, 4$.

\begin{itemize}
\item  Two more features are taken from Miller \cite{Miller96}: the two
ratios $T_{r}=\left\langle SuBr\right\rangle /I_{0}$ and $T_{cA}=I_{0}/\sqrt{%
A}$.

\item  Finally, five features from FOCAS \cite{Jarvis81}: the second and the
fourth total moments of the light distribution, the central intensity
averaged in a $3\times 3$ $area$, the ellipticity averaged over the whole
object area and, finally, the ''Kron'' radius defined as: 
\[
r_{Kron}=\frac{1}{\sum_{\left( x,y\right) \in A}I\left( x,y\right) }%
\sum_{\left( x,y\right) \in A}I\left( x,y\right) r\left( x,y\right) 
\]
\end{itemize}

\noindent 
In order to optimize the classification system performance it is necessary
to reduce the feature number. To do so we need training and test sets for a
subset of our objects. In our case we selected a subset of the Infante and
Pritchet catalog \cite{infante92}, \cite{Pritchet} built with deeper images
obtained under sub-arcsec seeing conditions. We experimented both
unsupervised and supervised NN's for both the feature selection and the
classification phases, but since unsupervised NN's did not reach appreciable
results, in this paper we present only result with MLP's.
\\
\\
\noindent 
The sequential backward elimination strategy \cite{Bishop95} works as
follows: let us suppose to initially have all $M$ features in one set and to
run the NN's with this set. Then, we build $M$ different sets with $M-1$
features each one and we run one NN for each set and take the set obtaining
the best classification, in this way eliminating the less significant
feature. Usually, after this first step the classification error decreases
if there are noisy or redundant features. Then, we repeat these steps
eliminating one feature at each step.
\\
\\
\noindent 
For what concerns supervised learning NN's, we used some MLP's \cite
{Bishop95} with one hidden layer of $20$ neurons and only one output,
assuming value $0$ for star and value $1$ for galaxy. After the training, we
calculate the NN output as $1$ if it is greater than $0.5$ and $0$ otherwise
for each pattern of the test set. The most performing learning algorithm was
a hybrid conjugate gradients-quasi Newton method to take advantage of both
the algorithms.
\\

\begin{center}
{\bf \large 4. Experimental Results}
\end{center}

\noindent
{\bf 4.1 The data}

\medskip
\noindent 
In order to test the performances of our method we used a 2000x2000 arcsec$^{2}$ 
area centered on the North Galactic Pole extracted from 
the slighly compressed POSS-II F plate n. 443 (available via network at the CADC).
POSS-II data were linearized using the sensitometric spots recorded on the
plate. The average FWHM\ of our data was 3 arcsec.
The same area has  been widely studied by others and, in particular, by
\cite{infante92}, \cite{Pritchet} who used deep observation obtained at the 
3.6 m CFHT telescope in the $F$\ photographic band under good seeing conditions 
(FWHM $<1$ arcsec), to derive a catalogue of objects
complete down to $m_{F}\sim 23$.
Their catalogue is therefore based on data of much better quality and
accuracy than ours.
\\
\\
\noindent 
The selected region, a relatively empty one, slightly penalizes our NN
detection algorithms which easily recognize objects of quite different sizes
and - on the contrary of what happens to other algorithms -
work well even on very crowded area, such as the center of nearby
clusters of galaxies, as our preliminary test on a portion of the Coma
clusters (imaged on the same POSS-II plate) shows \cite{wirn98}.
\\
\\
\noindent
{\bf 4.2 The processing}

\medskip
\noindent
This POSS-II\ field was processed through several NN detection
algorithms (PCA NN's, Hierrchical Unsupervised NN's, MLP's) and also through
S-Extractor (=SE\-x; \cite{Bertin96}) which has come to be a standard in the
astronomical community. For what the SEx application to our dataset
is concerned refer to 
\cite{Andreon99}.
\\
\\
\noindent 
For the NN's, we used the PCA NN's to reduce the input space to 3 dimensions.
Then we run the unsupervised NN's on the $3$-D input related to the $5\times
5$ and $3\times 3$ running windows (in our experiments the best performing
NN's were: Neural gas (NG3), ML-Neural gas (MLNG3 or MLNG5), ML-SOM (K5),
GCS+ML-Neural gas (NGCS5). We just wish to stress here that, since the
background subtraction is a vital part of the detection, and in order not to
give an unfair advantage to any of the detections algorithms, all algorithms
including SEx, were run on the same background subtracted image.
\\
\\
\noindent 
Fig. 1 gives the number of ``True'' objects detected by SEx (upper panel), 
\textit{id est} objects having a counterpart in the \cite{Pritchet} catalog. As it can
be seen, the SEx catalog is uncomplete for \ $m_{F}<21$ mag, 
which is roughly the plate completness limit. 
The lower panel shows instead the relative
performance of the NN's, defined as the ratio between the number of ``True'' objects
detected by the specific NN and SEx, respectively. 
All the NN's and SEx turn out to be roughly equivalent in detecting ``True'' objects 
brighter than $m_{F}=21$, while for objects fainter than the completeness limit of the plate, only
MLNG5 is as efficient as SEx, followed by MLNG3. Therefore, differences
among catalogs concern only galaxies fainter than the plate completeness
limit.
\\
\\
\noindent 
Fig. 2 shows the number of ``False'' objects detected by SEx (upper panel),
where ``False'' means objects not having a counterpart in the \cite
{infante92}\ catalog, and therefore include a few ``True'' objects not
catalogued by \cite{infante92}\ (mainly because they are too bright).
We believe that all objects brighter than $m_{F}=20$ mag are really ``True''
since they are detected both by SEx and NN's with high significance. 
The lower panel shows the relative performances of the NN's, defined as the 
ratio of the number of ``False'' objects detected by the NN and by SEx. For
objects brighter than $m_{F}=19$ mag, NN's and SEx have similar performances,
while at $m_{F}=19.7$ mag, SEx works better (but only for a few objects, see
upper panel). NN's catalogues present, however, less false detections. MLNG5,
which is also quite efficient in detecting ``True'' objects, has a $20\%\;$%
cleaner detection rate in the highly populous bin $m_{F}=21.7$ mag. MLNG3 is
less efficient in detecting ``True'' objects but is even cleaner of false
detections.
\\
\\
\noindent 
Fig. 3 shows the number of missed objects by SEx (upper panel). ``Missed''
means being in the \cite{infante92} catalog, but not included in our
catalogs. Obviously, the step increase below $21$ mag coincides with the completeness limit of our photographic material. The lower panel gives the relative performances of the NN's,
defined as the ratio between the number of objects missed by the specific 
NN and by SEx.
MLNG3 and MLNG5 have performances almost constant at $\sim 1$ mag, while the
other NN's miss objects at $m_{F}\sim 21-22$ mag which, however, are still
fainter than the plate completeness limit.
\\
\\
\noindent 
The class of ``Missed'' objects needs more attention. It is likely that most of
the objects fainter than $m_{F}=21$ mag are too faint to be detected with a $100\%$
confidence level, so we focus first on brighter objects. They can be divided in:

-- objects detected by \cite{infante92} which correspond to empty regions in
our images. They can be missing because variable, fast moving, or with an
overestimated luminosity in \cite{infante92}. They can also be missed
because spurious in the template catalog or simply because they are too faint.

-- ``True'', nearby objects which are blended in our image but not in that of
\cite{infante92};

-- parts of isolated single large objects incorrectly split by \cite{infante92};

-- a few detections aligned in the E-W direction on the two sides of
the images of a bright star. They are likely false objects (diffraction
spikes detected as individual objects). 
\\
\\
\noindent 
Therefore, a fair fraction of the ``Missed'' objects are truly non existent 
and the performances of our detection tools are therefore lower bounded
at $m_{F}<21$ mag. We wish to stress here that even though there is nothing like a perfect 
catalogue, the template by \cite{infante92} is among the best ones ever produced to our best knowledge. 
\\
\\
\noindent 
In \cite{infante92}, objects are classified in 2 major classes, star \& galaxies, 
and a few minor classes (merged, noise, spike, defects, etc.), that we neglect. 
The efficiency of the detection is shown in Fig.4 for three representative detection 
algoritms: MLNG5, K5, and SEx. 
At $m_F < 21$ mag, the detection efficiency is large, close to 1
and independent on the central concentration of the light. 
Please note that there are no objects in the image 
having $m_F <16$ mag and that in the following bin there are only 4 galaxies. 
At fainter magnitudes ($\sim 22-23$ mag) detection efficiencies
differ as a function of both the algorithm and of the light concentration. 
In fact, SEx, MLNG5, and to less extent K5, turn out to be more efficient 
in detecting galaxies rather than stars (in other words: ``Missed'' objects are 
preferentially stars). 
For SEx, a possible explanation is that a minimal area above the background is 
required in order for the object to be detected.   
At $m_F \sim 22-23$ mag, noise fluctuations can affect the isophotal area
of unresolved objects bringing it below the assumed treshold (4 pixels). 
This bias is minimum among the three considered detection algoritms, for the K5 NN. 
However, this is more likely due to the fact that K5 misses more galaxies than the other
algorithms, rather than to the fact that it detects more stars.
\\

\begin{figure}[tbp]
\vbox{
\hbox{
\psfig{figure=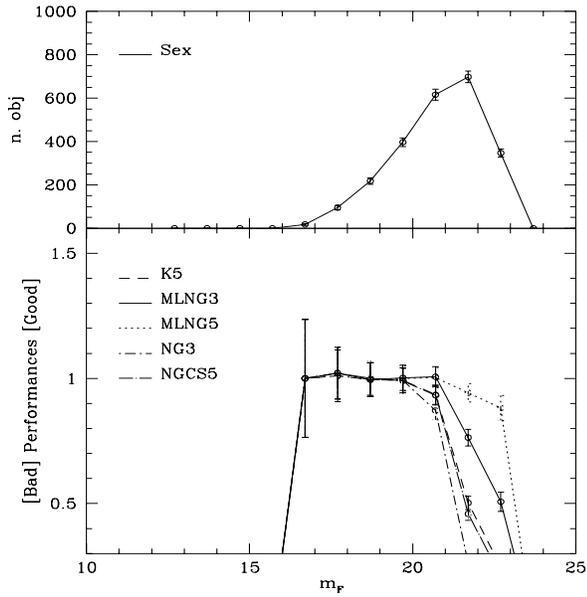,width=8cm,height=8cm,bbllx=40pt,bblly=195pt,bburx=472pt,bbury=682pt}
}}
\caption{Number of ``True" objects detected by SEx (upper panel); relative
performance of the NN's, defined as the ratio of the number of true objects
detected by the NN and by SEx, respectively (lower panel).
}
\end{figure}

\begin{figure}[tbp]
\vbox{
\hbox{
\psfig{figure=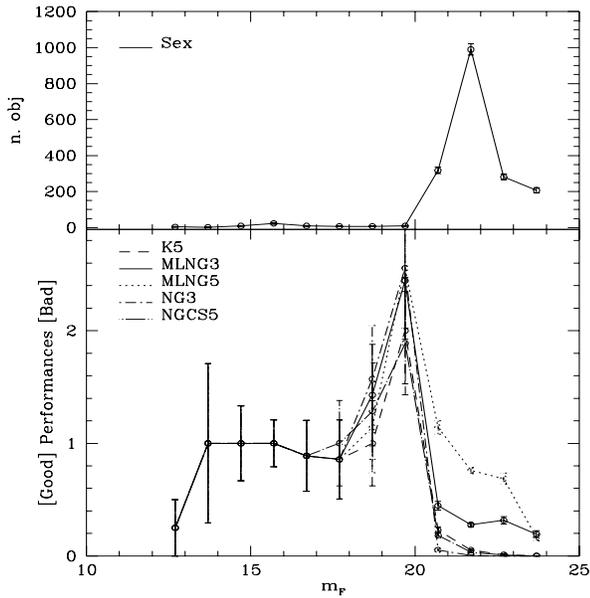,width=8cm,height=8cm,bbllx=44pt,bblly=196pt,bburx=473pt,bbury=681pt}
}}
\caption{Number of false objects detected by SEx (upper panel); 
relative performance of the NNs, defined as the ratio of the number of
``False'' objects detected by the NN and by SEx (lower panel).
}
\end{figure}

\begin{figure}[tbp]
\vbox{\hbox{
\psfig{figure=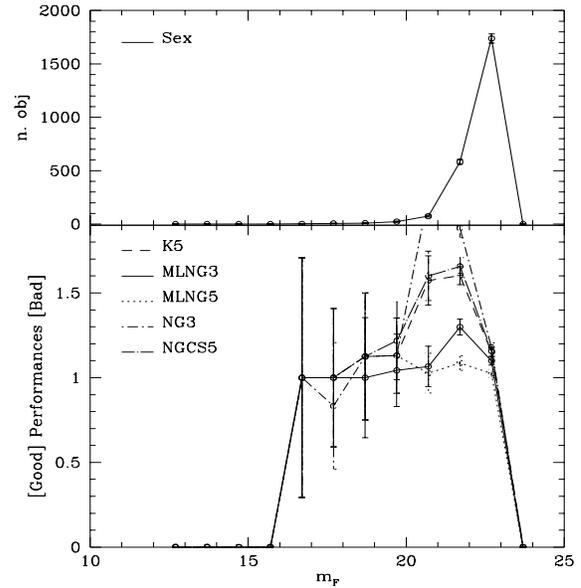,width=8cm,height=8cm,bbllx=35pt,bblly=190pt,bburx=481pt,bbury=682pt}
}}
\caption{Number of objects ``Missed" by SEx (upper panel); 
relative performance of the NN's, 
defined as the ratio of the number of objects missed by the NN and by SEx 
(lower panel).
}
\end{figure}

\begin{figure}[tbp]
\vbox{\hbox{
\psfig{figure=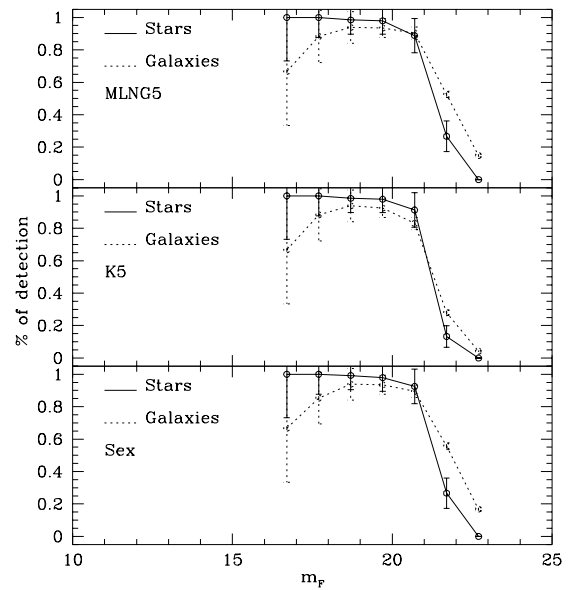,width=8cm,height=8cm,bbllx=45pt,bblly=196pt,bburx=486pt,bbury=640pt}
}}
\caption{Percent number of detected objects by MLNG5, K5 and SEx.
}
\end{figure}

\begin{center}
{\bf \large 5. Concluding Remarks}
\end{center}

\noindent 
In conclusion: MLNG3 and MLNG5 turn out to have performances
similar to SEx in detecting objects:\ they produce catalogs which are
cleaner of false detections but, at the same time, are also slightly more
uncomplete than SEx.
\\
\\
\noindent 
We also want to stress that since the less performing NN's produce
catalogs which are much cleaner of false detections, they can 
be used to select candidates for possible follow--up detailed studies 
at magnitudes where many of the objects detected by SEx would be false 
(i.e. the selected objects would be in large part true, and not just
noise fluctuations).

\noindent 
A posteriori, one could argue that performances similar to those of each of
the NN's could be achieved by running SEx with appropriate settings. However,
it would be unfair (and methodologically wrong) to make a fine tuning of any 
of the detection algorithms using a posteriori knowledge. 
\\

%
%
%

\end{document}